\documentclass[aps,physrev,twocolumn,preprint,floatfix,longbibliography,10pt]{revtex4-2}
\makeatletter
\def\l@subsubsection#1#2{}
\makeatother
\usepackage[english]{babel}
\usepackage[utf8]{inputenc}
\usepackage[T1]{fontenc}
\usepackage{microtype}
\usepackage{amsmath,amssymb,braket,slashed,mathrsfs}
\usepackage{pifont}
\usepackage{upgreek}

\usepackage[squaren,Gray,cdot,binary]{SIunits}
\usepackage{graphicx}
\graphicspath{{./figures/}}
\usepackage{color,hyperref}
\usepackage{minibox}
\usepackage{enumitem}
\usepackage{cleveref}
\usepackage{marginnote}

\crefformat{section}{Sec.~#2#1#3}
\crefmultiformat{section}{Secs.~#2#1#3}
{ and~#2#1#3}{, #2#1#3}{ and~#2#1#3}
\crefrangeformat{section}{Secs.~#3#1#4\,--\,#5#2#6}
\crefformat{subsection}{Sec.~#2#1#3}
\crefmultiformat{subsection}{Secs.~#2#1#3}
{ and~#2#1#3}{, #2#1#3}{ and~#2#1#3}
\crefrangeformat{subsection}{Secs.~#3#1#4\,--\,#5#2#6}
\crefformat{subsubsection}{Sec.~#2#1#3}
\crefmultiformat{subsubsection}{Secs.~#2#1#3}
{ and~#2#1#3}{, #2#1#3}{ and~#2#1#3}
\crefrangeformat{subsubsection}{Secs.~#3#1#4\,--\,#5#2#6}
\crefformat{figure}{Fig.~#2#1#3}
\crefmultiformat{figure}{Figs.~#2#1#3}
{ and~#2#1#3}{, #2#1#3}{ and~#2#1#3}
\crefrangeformat{figure}{Figs.~#3#1#4\,--\,#5#2#6}
\crefformat{table}{Table~#2#1#3}
\crefmultiformat{table}{Tables~#2#1#3}
{ and~#2#1#3}{, #2#1#3}{ and~#2#1#3}
\crefrangeformat{table}{Tables~#3#1#4\,--\,#5#2#6}
\crefformat{equation}{Eq.~(#2#1#3)}
\crefmultiformat{equation}{Eqs.~(#2#1#3)}{ and~(#2#1#3)}{, (#2#1#3)}{ and~(#2#1#3)}
\crefrangeformat{equation}{Eqs.~(#3#1#4)--(#5#2#6)}
\crefformat{appendix}{App.~#2#1#3}
\crefmultiformat{appendix}{Apps.~#2#1#3}
{ and~#2#1#3}{, #2#1#3}{ and~#2#1#3}
\crefrangeformat{appendix}{Apps.~#3#1#4\,--\,#5#2#6}

\usepackage{comment}

\definecolor{linkcolor}{rgb}{.17578125,.1875,.5703125}
\hypersetup{
  pdfstartview={FitH},
  pdftitle={TITLE},
  pdfauthor={AUTHOR},
  pdfsubject={Physics article},
  pdfcreator={Antonin Portelli},
  pdfkeywords={KEW1} {KEW2},
  colorlinks=true,
  linkcolor=linkcolor,
  citecolor=linkcolor,
  filecolor=black,
  urlcolor=linkcolor
}

\newcommand{\ie}{i.e.}

\DeclareMathOperator{\diag}{diag}

\newcommand{\Diff}{\mathrm{D}}
\renewcommand{\epsilon}{\varepsilon}
\newcommand{\mat}[1]{
  \begin{pmatrix}#1
\end{pmatrix}}

\DeclareMathOperator{\SU}{SU}
\DeclareMathOperator{\tr}{tr}
\DeclareMathOperator{\U}{U}

\newcommand{\nn}{\nonumber}

\newcommand{\tm}{T}
\newcommand{\Gw}[1]{\Gamma_{1}\cdot #1}
\newcommand{\fullprop}{\raisebox{-0ex}{\includegraphics{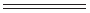}}}
\newcommand{\prop}{\raisebox{0.1ex}{\includegraphics{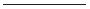}}}
\newcommand{\fulldminus}{\raisebox{-0.3ex}{\includegraphics{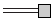}}}
\newcommand{\dminus}{\raisebox{-0.3ex}{\includegraphics{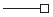}}}
\newcommand{\vcurrent}{\raisebox{-0.4ex}{\includegraphics{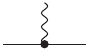}}}
\newcommand{\dvcurrent}{\raisebox{-0.4ex}{\includegraphics{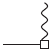}}}
\newcommand{\fullovprop}{\raisebox{-0.3ex}{\includegraphics{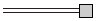}}}
\newcommand{\ovprop}{\raisebox{-0.3ex}{\includegraphics{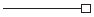}}}
\newcommand{\eOnej}{\raisebox{-0.3ex}{\includegraphics{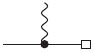}}}
\newcommand{\eOnedj}{\raisebox{-0.3ex}{\includegraphics{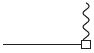}}}
\newcommand{\tcurrent}{\raisebox{-0.4ex}{\includegraphics{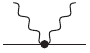}}}
\newcommand{\dtcurrent}{\raisebox{-0.4ex}{\includegraphics{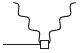}}}
\newcommand{\eTwojj}{\raisebox{-0.4ex}{\includegraphics{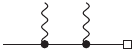}}}
\newcommand{\eTwojdj}{\raisebox{-0.4ex}{\includegraphics{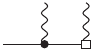}}}
\newcommand{\eTwot}{\raisebox{-0.4ex}{\includegraphics{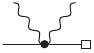}}}
\newcommand{\eTwodt}{\raisebox{-0.4ex}{\includegraphics{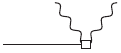}}}
\newcommand{\cern}{Theoretical Physics Department, CERN, 1211 Geneva 23, Switzerland}
\newcommand{\uoe}{School of Physics and Astronomy,
The University of Edinburgh, Edinburgh EH9 3FD, United Kingdom}

\begin{document}
\title{Perturbative quantum electrodynamics\texorpdfstring{\\}{}
with generalized domain wall fermions}
\preprint{CERN-TH-2026-172}
\author{M. Di Carlo}\affiliation{\cern}
\author{A. Portelli}\affiliation{\uoe}
\begin{abstract}
  In this paper we derive the expansion of the generalized domain-wall fermion Dirac
  operator including electromagnetic corrections up to $\mathcal{O}(e^2)$, which are
  relevant for lattice computations of radiative corrections to hadronic processes with
  chiral fermions. In the generalized formulation of the domain-wall fermionic QCD+QED
  action, physical quark fields are related to the corresponding five-dimensional fields
  in a way which depends on the (QCD+QED) gauge links, generating extra contact terms when
  expanding correlation functions with respect to the electric charge. We re-derive the 
  known first-order correction using a background-field approach and, at second order, 
  obtain new local operator insertions (seagull vertices) required for gauge covariant 
  calculations.
\end{abstract}
\maketitle
\section{Introduction}

In recent years, the inclusion of isospin-breaking effects and electromagnetic corrections in lattice QCD calculations has become increasingly important to achieve high-precision predictions for a number of hadronic observables. These effects arise from both the mass difference between up and down quarks and electromagnetic interactions, and are expected to play a significant role when precision reaches the sub-percent level. As the precision of experimental measurements increases, it becomes essential to account for these contributions in theoretical lattice calculations to ensure accurate comparisons.

To incorporate electromagnetic interactions on the lattice, two main approaches have been pursued. On the one hand, full non-perturbative QCD+QED simulations can be performed by generating gauge ensembles directly with active electromagnetic couplings and non-degenerate light quark masses (see, e.g., Refs.~\cite{BMW:2014pzb,RCstar:2022yjz}). On the other hand, since electromagnetic and strong isospin-breaking effects are small ($\alpha_{\mathrm{em}} \sim (m_d - m_u)/\Lambda_{\mathrm{QCD}} \sim 1\%$), a perturbative approach known as the RM123 method~\cite{deDivitiis:2013xla} has been widely adopted. This method consists in expanding path-integral correlation functions in powers of the electric charge $e$ and the quark mass difference around an isosymmetric QCD background ($e=0$), thus allowing one to reuse existing isosymmetric QCD gauge ensembles. We refer to~Ref.~\cite{Boyle:2017gzv,RC:2025zoa} for comparisons of the two approaches.
The RM123 method has been successfully applied to a broad spectrum of hadronic observables, including hadron mass splittings and light quark mass determinations~\cite{Giusti:2017dmp,Boyle:2017gzv,Frezzotti:2022dwn}, leptonic decay rates~\cite{Carrasco:2015xwa,Giusti:2017dwk,DiCarlo:2019thl,Boyle:2022lsi}, and the hadronic contributions to the anomalous magnetic moment of the muon~\cite{Giusti:2019xct,Djukanovic:2024cmq,RBC:2024fic}.

A general consequence of the RM123 perturbative expansions is that, at each order in $e$, additional operator insertions appear in the lattice correlation functions.
These insertions can be interpreted as effective vertices induced by the discretized action. Although they have no direct continuum analogue, they are required to preserve gauge invariance order by order at fixed lattice spacing and, consequently, the renormalizability of the lattice theory. Their explicit form depends on the discretization chosen for the fermionic action. 
At first order in $e$, the operator insertion corresponds to the conserved electromagnetic current, while at $\mathcal{O}(e^2)$ the new operator insertions are often referred to as seagull vertices.
While these terms are well established for Wilson-like fermions~\cite{deDivitiis:2013xla} and for the Shamir implementation of domain-wall fermions (DWF)~\cite{Boyle:2017gzv}, an explicit derivation of the seagull vertex for generalized DWF has been missing. As a result, several studies involving generalized DWF have neglected these contributions and replaced the conserved electromagnetic current with the corresponding local current, multiplied by an appropriate renormalization factor. Although this approach is correct when using a consistent renormalization procedure, the associated lattice theory is not $\U(1)$ gauge invariant, and potential consequences need to be reviewed on an observable by observable basis.
In this work we provide an explicit derivation of the electromagnetic expansion of the generalized domain-wall fermion Dirac operator up to $\mathcal{O}(e^2)$, including the previously missing local seagull vertices. We re-derive the first-order result using a background-field approach and present the new operator insertions arising at second order. These originate from the fact that, beyond the Shamir formulation, the relation between physical quark fields and the underlying five-dimensional fields depends explicitly on the gauge links, leading to additional contact interactions in the expansion in the electromagnetic coupling.

The paper is organized as follows. In Section~\ref{sec:definitions} we introduce the notation and review the generalized DWF setup. Section~\ref{sec:order1} revisits the $\mathcal{O}(e)$ expansion using the background-field method, while Section~\ref{sec:order2} contains the main derivation of the $\mathcal{O}(e^2)$ seagull vertices and a discussion of their properties. In Section~\ref{sec:orderN} we comment on the structure of the expansion at higher orders. Section~\ref{sec:qnumber} illustrates the impact of these terms on quark-number susceptibilities, and Section~\ref{sec:numerical} outlines practical aspects of the numerical implementation. Finally, we summarize our results and discuss future prospects in Section~\ref{sec:conclusion}.

\section{Generalized domain-wall fermions}
\label{sec:definitions}

We consider the generalized form of the domain-wall fermion (DWF) action introduced in
Refs.~\cite{Brower:2004xi,Brower:2005qw,Brower:2012vk}. The generalized DWF action can be defined in
terms of a 5-dimensional Dirac operator $D_\mathrm{dw}$ as
\begin{equation}
  \mathcal{S}_\mathrm{DWF}= \sum_{x\in\Lambda} \bar{\psi}(x) (D_\mathrm{dw}\psi)(x)\,,
  \label{eq:dwfaction}
\end{equation}
where $x$ denotes the physical 4-dimensional spacetime coordinate, and $\psi(x)$ and
$\bar{\psi}(x)$ are 5-dimensional fermion and anti-fermion fields, respectively. Here the fifth dimension is considered finite with $L_s$ sites indexed from $1$ to $L_s$, and $\psi$ is interpreted as a column vector
of 4-dimensional fields, \ie
\begin{equation}
  \psi(x)^T=(\psi_1(x),\dots,\psi_{L_s}(x))\,.
\end{equation}
The Dirac operator $D_\mathrm{dw}$, which is a function of the Lagrangian bare quark mass
$m$, is defined by the following matrix in the fifth dimension
\begin{widetext}
  \begin{equation}
    \label{eq:gen_dirac_op_Brower}
   D_\mathrm{dw}(m)=\left(\begin{array}{c*{5}{@{\hskip 1pt}c@{\hskip 1pt}}}
      D_+^1           & -D_-^1P_-    & 0      & \cdots & 0           & m\,D_-^1P_+ \\
      -D_-^2P_+        & \ddots    & \ddots & \ddots & \vdots      & 0          \\
      0             & \ddots    & \ddots & \ddots & 0           & \vdots     \\
      \vdots        & \ddots    & \ddots & \ddots & \ddots      & 0          \\
      0             & \cdots    & 0      & \ddots & \ddots      & -D_-^{L_s-1}P_- \\
      m\,D_-^{L_s}P_-& 0         & \cdots & 0      & -D_-^{L_s}P_+  & D_+^{L_s}
        \end{array}\right)\,,
  \end{equation}
\end{widetext}
where the operators $D_{\pm}^s$ are given by
\begin{equation}
  D_+^s=1+b_s \, D_\mathrm{w}\,,\qquad
  D_-^s=1-c_s \, D_\mathrm{w}\,.
  \label{eq:dirac_op_plus_minus}
\end{equation}
The coefficients $b_s$ and $c_s$ are in general complex numbers, while $D_\mathrm{w}$ is
the usual Wilson-Dirac operator with (negative) mass $M$ acting on the physical
dimensions,
\begin{equation}
  D_\mathrm{w}=4+M-
  \sum_{\mu}(\Gamma_{\mu}^++\Gamma_{\mu}^-)\,.
  \label{eq:DW}
\end{equation}
The forward and backward Wilson kernels are defined respectively as
\begin{align}
  \Gamma_{\mu}^+(x,y)&=\frac12\,(1-\gamma_\mu)\,
    U_\mu(x)\,\delta(x-y+\hat{\mu})\,,\\
  \Gamma_{\mu}^-(x,y)&=\frac12\,(1+\gamma_\mu)\,
  \,U_\mu^\dagger(y)\,\delta(x-y-\hat{\mu})\,,
\end{align}
where $U_{\mu}$ are $\SU(3)$ QCD gauge links and $\gamma_\mu$ are the Euclidean gamma matrices satisfying the anticommutation relations $\{\gamma_\mu, \gamma_\nu\} = 2\delta_{\mu\nu}$.

We define the diagonal matrices
\begin{equation}
  D_\pm = \mathrm{diag}(D_\pm^1, \dots, D_\pm^{L_s})\,,
\end{equation}
and introduce the matrix
\begin{equation}
  \mathcal{P} =
  \mat{
    P_{-}  & P_{+}  & 0      & \cdots & 0      & 0      \\
    0      & P_{-}  & P_{+}  & 0      & \cdots & 0      \\
    0      & 0      & P_{-}  & P_{+}  & 0      & \vdots \\
    \vdots & 0      & \ddots & \ddots & \ddots & 0      \\
    0      & \cdots & 0      & \ddots & P_{-}  & P_{+}  \\
    P_{+}  & 0      & \cdots & 0      & 0      &  P_{-}
  }\,.
\end{equation}
Both $D_\pm$ and $\mathcal{P}$ act on the fifth-dimensional and spin indices.
The matrix $\mathcal{P}$ is unitary and $P_{\pm}$ denote the chiral projectors
\begin{equation}
  P_\pm=\frac12(1\pm\gamma_5)\,,
\end{equation}
which satisfy the orthogonal projector identities $P_{\pm}^2=P_{\pm}$, $P_++P_-=1$, and $P_{+}P_{-}=0$.
Using these matrices, the Dirac operator in~\cref{eq:gen_dirac_op_Brower} can be expressed as
\begin{equation}
  D_\mathrm{dw}(m) = [D_+ P_- + D_- P_+] \, D_\chi(m)\, \mathcal{P}^{-1} \,.
\end{equation}
Starting from this expression, it is possible to
show~\cite{Brower:2005qw,Brower:2012vk,RBC:2014ntl} that the Schur complement of the
matrix $D_\chi(m)$ with respect to the component $s=s'=1$ is given by the 4-dimensional
operator
  \begin{equation}
    S_\chi(m) = (1+\tm^{-L_s})^{-1} \, \gamma_5\,
    \bigg[
      \frac{1+m}{2} + \frac{1-m}{2}  \, \gamma_5 \, \frac{1-\tm^{-L_s}}{1+\tm^{-L_s}}
    \bigg] \,,
  \end{equation}
where we have used the shorthand notation
\begin{equation}
  \tm^{-L_s} = T_1^{-1}T_2^{-1}\cdots T_{L_s}^{-1} \,,
\end{equation}
and defined the inverse transfer matrix as
\begin{equation}
  T^{-1}_s = [1-{H}_s]^{-1}\,[1+{H}_s]\,,
\end{equation}
with the kernel function
\begin{equation}
  {H}_s = \gamma_5 \, \frac{(b_s+c_s)\,D_\mathrm{w}}{2+(b_s-c_s)D_\mathrm{w}} \,.
  \label{eq:kernel_function}
\end{equation}

An approximation of the overlap Dirac operator is then given by
\begin{align}
  D_\mathrm{ov}(m)&\simeq  S_\chi(1)^{-1} S_\chi(m) \nn\\
  &= \frac{1+m}{2} + \frac{1-m}{2}  \, \gamma_5 \, \frac{1-\tm^{-L_s}}{1+\tm^{-L_s}}\,.
  \label{eq:ov_approx}
\end{align}
In the limit $L_s \to \infty$, the ratio $(1 - \tm^{-L_s})/(1 + \tm^{-L_s})$ approaches
the sign function $\epsilon(\log \tm)$.
Different choices of $b_s$ and $c_s$ lead to different approximations of the overlap Dirac
operator, which are all equivalent in the limit ${L_s\to\infty}$ and discussed below in~\cref{sec:numerical}.

Furthermore, it can be shown (see e.g. Refs.~\cite{Brower:2012vk,RBC:2014ntl}) that
\begin{align}
  D_\mathrm{ov}(m) &\simeq S_\chi(1)^{-1} S_\chi(m) \nn\\
  & = \big[D_\chi(1)^{-1}D_\chi(m)\big]_{11} \nn\\
  &= \big[\mathcal{P}^{-1}D_\mathrm{dw}(1)^{-1}D_\mathrm{dw}(m)\mathcal{P}\big]_{11}\;.
  \label{eq:ov_dwf_correspondence}
\end{align}
Following Refs.~\cite{Narayanan:1994gw,Edwards:1998wx}, the physical quark propagator is
obtained from the massive overlap operator as
\begin{equation}
  S_q = \frac{1}{1-m}\,\big[D_\mathrm{ov}^{-1}-1\big]\,.
  \label{eq:ovprop_def}
\end{equation}
This relation comes from the fact that the massless overlap operator satisfies the
Ginsparg-Wilson relation
\begin{equation}
  \{D_\mathrm{ov}(0)^{-1},\gamma_5\} = 2\gamma_5\,.
\end{equation}
As a result, $D_\mathrm{ov}(0)^{-1}$ contains a contact term proportional to the identity,
which is removed by the subtraction in~\cref{eq:ovprop_def}. The overall factor
$(1-m)^{-1}$ comes from the mass rescaling of the massive overlap operator, $D_\mathrm{ov}
= (1-m) D_\mathrm{ov}(0) + m$, and ensures the correct normalization of the propagator.

Therefore, we obtain
\begin{align}
  S_q &=\frac{1}{1-m}\,[\mathcal{P}^{-1} \, S_\mathrm{dw} D_\mathrm{dw}(1) \,
  \mathcal{P}-1]_{11} \\
  &= [\mathcal{P}^{-1} \, S_\mathrm{dw} \,D_-\, \mathcal{R}_5 \mathcal{P}]_{11}\,,
  \label{eq:ovprop}
\end{align}
where $S_\mathrm{dw} = D_\mathrm{dw}(m)^{-1}$, and we used the relation
\begin{equation}
  \big[D_\mathrm{dw}(1) - D_\mathrm{dw}(m)\big] = (1-m) D_- \mathcal{R}_5\;,
  \label{eq:Ddw_diff}
\end{equation}
with
\begin{equation}
  (\mathcal{R}_5)_{s,s'} \equiv \delta_{s,L_s}\delta_{s',1} \, P_{-} +
  \delta_{s,1}\delta_{s',L_s}\, P_{+} \, .
\end{equation}

Equation~\eqref{eq:ovprop} implicitly defines the physical quark fields as
\begin{align}
  q(x) &= [\mathcal{P}^{-1}\psi(x)]_1 = P_- \, \psi_1(x) + P_+ \, \psi_{L_s}(x)\\
  \bar{q}(x) &= [(\bar{\psi}D_{-})(x)\mathcal{R}_5\mathcal{P}]_1 \\
  &= (\bar{\psi}_1\, D_-^1)(x) \, P_+ + (\bar{\psi}_{L_s} \, D_-^{L_s})(x)\, P_-\,.
  \label{eq:phys_fields}
\end{align}
Thus, the left- and right-handed components are localized at the boundaries of the fifth
dimension, $s = 1$ and $s = L_s$. Note that, given our convention on the definition of the
operator $D_\mathrm{dw}$ in~\cref{eq:gen_dirac_op_Brower}, the physical antiquark field
$\bar{q}$ acquires a dependence on the gauge links $U_\mu$ through the operator $D_-$.
Additionally, we define the quark path integral
\begin{equation}
  \braket{O}=\frac{1}{\mathcal{Z}}\int\Diff\psi\,\Diff\bar{\psi}\,
  O[\psi,\bar{\psi}]\,e^{-\mathcal{S}_\mathrm{DWF}[\psi,\bar{\psi}]}\,,
  \label{eq:pint}
\end{equation}
where $\mathcal{Z}$ is such that $\braket{1}=1$. The propagators $S_\mathrm{dw}$ and
$S_q$ are then given by the expectation values
\begin{align}
  S_\mathrm{dw}(x,y)&=\braket{\psi(x)\bar{\psi}(y)}\,,\quad\text{and}\\
  S_q(x,y)&=\braket{q(x)\bar{q}(y)}\,.
\end{align}
One can couple quark fields to an arbitrary background electromagnetic potential $A_{\mu}$ by 
promoting the gauge links $U_\mu$ in~\cref{eq:DW} to $\U(3)$ links through
\begin{equation}
  \mathcal{U}_\mu=\exp(i e A_{\mu})\,U_{\mu}\,,
  \label{eq:u3links}
\end{equation}
where $e$ is the electromagnetic coupling.
In this paper we are mainly interested in the expansion of the physical quark propagator $S_q$
in~\cref{eq:ovprop} with respect to the electric charge $e$,
\begin{equation}
  S_q = (S_q)|_{e=0} + e \, (\partial_e S_q)|_{e=0} + \frac{1}{2}e^2 \,
  (\partial_e^2 S_q)|_{e=0} + \mathcal{O}(e^3)\,,
  \label{eq:sqexp}
\end{equation}
which is relevant for the computation of radiative corrections to hadronic observables.

\section{First-order corrections}
\label{sec:order1}
In this section we discuss the derivation of the first-order correction $(\partial_e
S_q)|_{e=0}$ in \cref{eq:sqexp}. In order to simplify the notation, unless otherwise stated,
all derivatives with respect to the electric charge will be considered evaluated at $e=0$.
\subsection{Vector current background field derivation}
We begin by observing that the derivative of the $\U(3)$ links at $e=0$ is given by
\begin{equation}
  \partial_e\,\mathcal{U}_\mu=iA_{\mu}\,U_\mu\,.
\end{equation}
Then, since the matrices $\mathcal{P}$ and $\mathcal{R}_5$ in~\cref{eq:ovprop} are
independent of the gauge field, we can focus on the expansion of the product of the
five-dimensional propagator $S_\mathrm{dw}$ and the operator $D_-$. We start from the
well-known first-order expansion in the electric charge $e$ of the Wilson-Dirac operator
in~\cref{eq:DW},
\begin{equation}
  \partial_e D_\mathrm{w}=i\,(\Gw{A})\,,
\end{equation}
where
\begin{equation}
  (\Gw{A})(x,y) = \sum_{\mu,z\in\Lambda} A_\mu(z) \, \Gamma_1^\mu(x,z,y)\,,
\end{equation}
and we have introduced the point-split operator
\begin{equation}
  \Gamma_1^\mu(x,z,y) = \Gamma_\mu^-(x,y) \, \delta(z-y) - \Gamma_\mu^+(x,y) \, \delta(z-x)\,.
\end{equation}
Note that $(\Gw{A})$ is the discrete version of the continuum local interaction kernel
$\slashed{A}=\sum_{\mu}A_{\mu}\gamma_{\mu}$. Here the subscript~$1$ indicates that
the operator $\Gamma_1^\mu$ comes from the first-order perturbation of the Dirac operator.
One can then easily derive the following relations
using~\cref{eq:dirac_op_plus_minus,eq:gen_dirac_op_Brower} above
\begin{align}
  \partial_e D_+^s   &=
  i\,b_s\,\Gw{A}\,, \nn \\
  \partial_e D_-^s   &=-i\,c_s\,\Gw{A}\,,
  \label{eq:op_1st_derivative}\\
  \partial_e D_\mathrm{dw}(m)&=
  i\,(\Gw{A})\,\Omega(m)\,, \nn
\end{align}
where $\Omega(m)$ is the matrix
  \begin{equation}
    \Omega(m)=\left(\begin{array}{c*{5}{@{\hskip 1pt}c@{\hskip 1pt}}}
      b_1           & c_1P_-    & 0      & \cdots & 0           & -m\,c_1P_+ \\
      c_2P_+        & \ddots    & \ddots & \ddots & \vdots      & 0          \\
      0             & \ddots    & \ddots & \ddots & 0           & \vdots     \\
      \vdots        & \ddots    & \ddots & \ddots & \ddots      & 0          \\
      0             & \cdots    & 0      & \ddots & \ddots      & c_{L_s-1}P_- \\
      -m\,c_{L_s}P_-& 0         & \cdots & 0      & c_{L_s}P_+  & b_{L_s}
        \end{array}\right)\,.
  \end{equation}
  Since $S_\mathrm{dw}=D_\mathrm{dw}^{-1}$, one has
  \begin{equation}
    \partial_eS_\mathrm{dw}=-S_\mathrm{dw}\left[\partial_eD_\mathrm{dw}(m)\right]S_\mathrm{dw}\,,
  \end{equation}
and defining the quantity $S=S_\mathrm{dw} D_-$ entering~\cref{eq:ovprop} one obtains
\begin{equation}
  \partial_e
  S=-S_\mathrm{dw}\left[\partial_eD_\mathrm{dw}(m)\right]S+S_\mathrm{dw}\left(\partial_eD_-\right)\,.
  \label{eq:deS}
\end{equation}
This last expression can be rewritten using~\cref{eq:op_1st_derivative} as
\begin{align}
  \partial_e S_\mathrm{dw}
  &=-i\,S_\mathrm{dw}(\Gw{A})\,\Omega(m)\,
  S_\mathrm{dw}\,,
  \label{eq:dworder1}\\
  \partial_e S
  &=-i\,S_\mathrm{dw}(\Gw{A})\,\left[\Omega(m)
  \,S +C \right]\,,
  \label{eq:order1}
\end{align}
with $C=\diag(c_1,\dots,c_{L_s})$. Setting $e=0$ gives the first-order correction to $S$,
from which the correction to the physical quark propagator $S_q$ is readily
obtained using~\cref{eq:ovprop},
\begin{equation}
  \partial_e S_q = [\mathcal{P}^{-1} \, (\partial_e S) \, \mathcal{R}_5
  \mathcal{P}]_{11}\,.
  \label{eq:order1_phi}
\end{equation}

Let us discuss the properties of this result. For standard Shamir domain-wall
fermions~\cite{Shamir:1993zy,Furman:1994ky}, one has $b_s=1$ and $c_s=0$, which directly implies
$S=S_\mathrm{dw}$, $\Omega(m)=1$, and $C=0$. With these simplifications we recover the usual
form
\begin{equation}
  \partial_eS\underset{\mathrm{Shamir}}{=}-iS\,(\Gw{A})\,S\,,
\end{equation}
which corresponds to the insertion of the conserved Wilson vector current
between two QCD five-dimensional propagators when evaluated at $e=0$. 
For non-zero values of the $c_s$ coefficients, one obtains instead the additional term
$-i\,\smash{S_\mathrm{dw}(\Gw{A})\,C}$, which cannot be interpreted as a current
insertion. This term comes from the QED corrections to the operator $D_-$ appearing 
in the definition of the antiquark field in~\cref{eq:phys_fields}, and contributing 
to the physical propagator in~\cref{eq:ovprop}, namely
\begin{equation}
  \partial_e\,\bar{q}=i(\bar{Q}_1\cdot A)\,,
\end{equation}
with
\begin{equation}
  (\bar{Q}_1\cdot A)(x) = \sum_{\mu,z\in\Lambda} A_\mu(z) \, \bar{Q}_1^\mu(z,x)\,,
\end{equation}
and having defined
\begin{equation}
  \bar{Q}_1^\mu(z,x)= -\sum_{x'\in\Lambda} \, \{\bar{\psi}(x')\,\Gamma_1^\mu(x',z,x)\,C\,\mathcal{R}_5\mathcal{P}\}_1\,.
\end{equation}
Note that, given the point-split nature of the operator $\Gamma_1^\mu$, the
derivative of the field $\partial_e \bar{q}(x)$ has support at $x$ and at its
nearest-neighbor sites.
With this definition, together with~\cref{eq:phys_fields}, we can rewrite the
first order correction to the physical propagator in~\cref{eq:order1_phi} as
\begin{align}
  \partial_e S_q(x,y) 
  &= - i\,\sum_{\mu,z\in\Lambda}A_\mu(z) \, \Big\{ \langle q(x) V^\mu(z) \bar{q}(y)  \rangle
   \nn\\
  &\qquad\qquad\qquad\qquad
   -\langle q(x) \bar{Q}_1^\mu(z,y) \rangle \Big\}\,,
\end{align}
where we have introduced the DWF conserved current
\begin{equation}
  V^{\mu}(z)=\sum_{x,y\in\Lambda} \bar\psi(x) \, \Gamma_1^\mu(x,z,y) \, \Omega(m) \psi(y)\,,
  \label{eq:dwf_conserved_current}
\end{equation}
which can be explicitly written as
\begin{align}
  V^\mu(z) &=-\frac{1}{2}[\bar{\psi}(z)(1-\gamma_{\mu})U_{\mu}(z)\Omega(m)\psi(z+\hat{\mu})\nn\\
  &\quad\qquad-\bar{\psi}(z+\hat{\mu})(1+\gamma_{\mu})U_{\mu}^\dagger(z)\Omega(m)\psi(z)
  ]\,.
\end{align}
Analogously, we can rewrite the first-order correction to the DWF quark propagator as
\begin{equation}
  \partial_e S_\mathrm{dw}(x,y) = - i\,\sum_{\mu,z\in\Lambda}A_\mu(z) \, \langle \psi(x) V^\mu(z) \bar{\psi}(y)  \rangle\,.
\end{equation}
In the next subsection we show that $V^\mu(z)$ is the conserved vector current obtained
from a local $\U(1)$ transformation of the quark fields in the DWF action and derive
the associated Ward identities.

One can conveniently represent the result in~\cref{eq:order1} using Feynman diagrams. We
start by defining the following position-space Feynman rules
\begin{alignat}{3}
  \fullprop   &= S_\mathrm{dw} && \qquad \prop    &&=(S_\mathrm{dw})|_{e=0}\\
  \fulldminus &= D_-          && \qquad \dminus  &&=(D_-)|_{e=0}\\
  &\vcurrent  &&= i\,(\Gw{A})\,\Omega(m) \hspace{-2cm}&&\label{eq:Vvertex}\\
  &\dvcurrent &&=-i\,(\Gw{A})\,C\,. \hspace{-2cm}&&
\end{alignat}

Then the first-order expansion of $S$ in~\cref{eq:order1} can be written as
\begin{widetext}
  \begin{equation}
    S= \fullovprop =
    \ovprop +e\,\Bigg( - \eOnej + \eOnedj~\Bigg)+\mathcal{O}(e^2)\,.
  \end{equation}
\end{widetext}

\subsection{Field transformations and Ward identities}
In the previous section, we derived the conserved $\U(1)$ vector current arising in
first-order electromagnetic corrections directly from the response of a quark propagator
to a background field. In this section, we discuss how the identities obtained are
equivalent to the traditional approach using a local $\U(1)$ transformation of the quark
fields.

We consider the $\U(1)$ vector transformations
\begin{equation}
  \psi\underset{V}{\mapsto}g\psi,\quad\text{and}\quad
  \bar{\psi}\underset{V}{\mapsto}g^*\bar{\psi}\,,
  \label{eq:quarkvtrans}
\end{equation}
where $g(x)=e^{ie\omega(x)}$ and $\omega(x)$ is an arbitrary real field.
We additionally define $\delta_V$ as the operator giving the variation under an infinitesimal vector transformation, \ie
\begin{equation}
  \delta_V\psi=ie\omega\psi,\quad\text{and}\quad\delta_V\bar{\psi}=-ie\omega\bar{\psi}\,.
\end{equation}
The vector transformation in \cref{eq:quarkvtrans} is unitary, and therefore can be absorbed
with a change of variables in the path integral in \cref{eq:pint}, with no change in the
integration measure. In summary, for any observable $O$, we can write
\begin{equation}
  \delta_V\langle{O}\rangle=0\,.
\end{equation}
Computing the variation under the path integral and applying the product rule, we obtain
\begin{equation}
  \braket{O\,\delta_V\mathcal{S}_\mathrm{DWF}}=\braket{\delta_V O}\,,
\end{equation}
which is a condensed form for the Ward-Takahashi identity. Let us compute the variation of
the DWF action.

Applying the transformation \eqref{eq:quarkvtrans} to the DWF
action in~\cref{eq:dwfaction}, one obtains
\begin{equation}
  \mathcal{S}_\mathrm{DWF}\underset{V}{\mapsto} \sum_{x\in\Lambda} 
  \bar{\psi}(x)(g^*D_\mathrm{dw}\,g\,\psi)(x)\,.
\end{equation}
Therefore, a vector transformation is equivalent to the Dirac operator transformation
\begin{equation}
  D_\mathrm{dw}(x,y)\underset{V}{\mapsto}g^*(x)D_\mathrm{dw}(x,y)g(y)\,.
  \label{eq:ddwvtrans}
\end{equation}
Now, defining the forward and backward finite-difference operators as
\begin{align}
  \partial_\mu f(x) = f(x+\hat{\mu}) - f(x)\,, \\
  \partial_\mu^* f(x) = f(x) - f(x-\hat{\mu})\,,
\end{align}
respectively, and using the definitions in~\cref{eq:DW,eq:dirac_op_plus_minus,eq:gen_dirac_op_Brower},
one can show that this transformation is equivalent to
\begin{equation}
  U_{\mu}\underset{V}{\mapsto}g^*U_{\mu}\,(\tau_{\mu}g)=e^{ie\partial_{\mu}\omega}\,U_{\mu}\,.
\end{equation}
In summary, the vector transformation \cref{eq:quarkvtrans} is equivalent to coupling to
the electromagnetic potential $\partial_{\mu}\omega$. From a gauge theory point of view,
this is expected as we know the theory becomes $\U(1)$ gauge-invariant if the
electromagnetic potential transforms as $A_{\mu}\mapsto A_{\mu}-\partial_{\mu}\omega$. In
particular, an infinitesimal vector transformation of the action can be obtained by
expanding around $e=0$ as in the previous section, \ie
\begin{equation}
  \delta_V\mathcal{S}_\mathrm{DWF}[U_{\mu}]=e\,
  \partial_e\mathcal{S}_\mathrm{DWF}[\mathcal{U}_{\mu}]|_{A_{\mu}
  =\partial_{\mu}\omega,e=0}\,,
\end{equation}
where we made the dependence on the $\SU(3)$ or $\U(3)$ gauge links explicit. The
expression above can now be computed using our previous results in~\cref{eq:op_1st_derivative}:
\begin{equation}
  \delta_V\mathcal{S}_\mathrm{DWF}=ie\sum_{x,y\in\Lambda}\bar{\psi}(x)\,(\Gw{\partial
  \omega})(x,y)\,\Omega(m)\,\psi(y)\,.
  \label{eq:actionvtrans}
\end{equation}
It is also instructive to derive this result directly. The infinitesimal version of the
transformation in~\cref{eq:ddwvtrans} can be written as the commutator
\begin{equation}
   D_\mathrm{dw}\underset{V}{\mapsto}D_\mathrm{dw} -ie[\omega,D_\mathrm{dw}] \,.
\end{equation}
where, in this context, $\omega$ is abusively understood as the identity operator multiplying
the field $\omega$. Additionally, one can show that
\begin{equation}
  [\omega,\Gamma_{\mu}^+]=-(\partial_{\mu}\omega)\,\Gamma_{\mu}^+
  \quad\text{and}\quad
  [\omega,\Gamma_{\mu}^-]=\Gamma_{\mu}^-\,(\partial_{\mu}\omega)\,,
\end{equation}
which implies
\begin{equation}
  [\omega,D_\mathrm{w}]=-(\Gw{\partial\omega})\,,
\end{equation}
and therefore
\begin{equation}
  [\omega,D_\mathrm{dw}]=-(\Gw{\partial\omega})\,\Omega(m)\,.
\end{equation}
This last equation directly implies \cref{eq:actionvtrans}. 
The action variation can also be written in terms of the DWF conserved vector current defined in~\cref{eq:dwf_conserved_current} as follows:
\begin{equation}
  \delta_V\mathcal{S}_\mathrm{DWF}=ie\sum_{\mu,x\in\Lambda}(\partial_{\mu}\omega)(x)V_{\mu}(x)\,.
\end{equation}
Then, using integration by parts, one gets
\begin{equation}
  \delta_V\mathcal{S}_\mathrm{DWF}=-ie\sum_{\mu,x\in\Lambda}\omega(x)\,\partial_{\mu}^*V_{\mu}(x)\,.
\end{equation}
With the particular choice $\omega(x)=\delta(x-z)$, we finally obtain the Ward-Takahashi
identity
\begin{equation}
  -ie\,{\textstyle\sum}_{\mu}\partial_{\mu,z}^*\braket{V_{\mu}(z)\,O}=
  \left.\braket{\delta_VO}\right|_{\omega(x)=\delta(x-z)}\,.
  \label{eq:wti}
\end{equation}

We can now derive the Ward-Takahashi identity for the electromagnetic vertices for both
domain-wall and physical quark fields. 
In the case of domain-wall quarks, choosing the operator ${O=\psi(x)\bar{\psi}(y)}$ in \cref{eq:wti}, one obtains
\begin{align}
  {\textstyle\sum}_{\mu}\partial_{\mu,z}^*\braket{\psi(x)V_{\mu}(z)\bar{\psi}(y)}&=
  \delta(y-z)S_\mathrm{dw}(x,y)\nn\\
  &\quad-\delta(x-z)S_\mathrm{dw}(x,y)\,,
\end{align}
which relates the five-dimensional electromagnetic vertex and the domain-wall propagators.
On the other hand, in the case of the physical four-dimensional fields, the Ward-Takahashi identity
contains an additional term when ${C\neq 0}$, arising from the operator $D_{-}$ in the definition of the
antiquark field. In fact, choosing $O=q(x)\bar{q}(y)$, one has
\begin{align}
  \delta_V\bar{q}(y)&=-ie[(\bar{\psi}\omega D_-)(y)\mathcal{R}_5\mathcal{P}]_1\\
  &=-ie\omega(y)\bar{q}(y)-ie[(\bar{\psi}[\omega,D_-])(y)\mathcal{R}_5\mathcal{P}]_1\\
  &=-ie\omega(y)\bar{q}(y)-ie\{[\bar{\psi}(\Gw{\partial\omega})](y)C\mathcal{R}_5\mathcal{P}\}_1\,.
\end{align}
Therefore, with some algebra, one obtains the identity
\begin{align}
  {\textstyle\sum}_{\mu}\partial_{\mu,z}^*\braket{q(x)V_{\mu}(z)\bar{q}(y)}&=
  \delta(y-z)S_q(x,y)\nn\\
  &-\delta(x-z)S_q(x,y)\nn\\
  &+{\textstyle\sum}_{\mu}\partial_{\mu,z}^*\braket{q(x)\bar{Q}_{1}^{\mu}(z,y)}\,.
\end{align}
In the expression above, the last term is an extended contact term which only contributes
when $z$ coincides with $y$ or its nearest-neighbor sites.

\section{Second-order corrections}
\label{sec:order2}
In this case we start from the second derivative of the $\U(3)$ links at $e=0$, which is given by
\begin{equation}
  \partial_e^2\,\mathcal{U}_\mu=-A_{\mu}^2\,U_\mu\,.
\end{equation}
The second-order expansion of the physical propagator $S_q$ is related to the
second-order derivative in $e$ of the Wilson-Dirac operator, namely
\begin{equation}
  \partial_e^2D_\mathrm{w}=(\Gamma_2\cdot A^2)\,,
\end{equation}
where
\begin{equation}
  (\Gamma_2\cdot A^2)(x,y) = \sum_{\mu,z\in\Lambda} A_\mu(z)^2 \, \Gamma_2^\mu(x,z,y)\,,
\end{equation}
and we have introduced the point-split operator appearing at second order as 
\begin{equation}
  \Gamma_2^\mu(x,z,y) = \Gamma_\mu^-(x,y) \, \delta(z-y) + \Gamma_\mu^+(x,y) \, \delta(z-x)\,.
\end{equation}
Combining this with~\cref{eq:dirac_op_plus_minus,eq:gen_dirac_op_Brower} leads directly
to the following second-order derivatives:
\begin{align}
  \partial_e^2D_+^s&=b_s\, \Gamma_2\cdot A^2 \,,\nn\\
  \partial_e^2D_-^s&=-c_s\, \Gamma_2\cdot A^2 \,,
  \label{eq:op_2nd_derivative}\\
  \partial_e^2D_\mathrm{dw}&= (\Gamma_2\cdot A^2) \, \Omega(m) \,.\nn
\end{align}
Differentiating~\cref{eq:deS} once more with respect to $e$ gives
\begin{align}
  \partial_e^2S&=2\,S_\mathrm{dw}[\partial_eD_\mathrm{dw}(m)]S_\mathrm{dw}[\partial_eD_\mathrm{dw}(m)]S\notag\\
  &-S_\mathrm{dw}[\partial_e^2D_\mathrm{dw}(m)]S-2S_\mathrm{dw}[\partial_eD_\mathrm{dw}(m)]S_\mathrm{dw}(\partial_e D_-)\notag\\
  &+S_\mathrm{dw}(\partial_e^2 D_-)\,,
\end{align}
which can be rewritten as
  \begin{align}
    \partial_e^2S&=-2\, S_\mathrm{dw}\,(\Gamma_1\cdot A) \, \Omega(m)\,
      S_\mathrm{dw} \, (\Gamma_1\cdot A) \left[\Omega(m) \, S+C\right]\nn\\
    &\quad -S_\mathrm{dw}\,(\Gamma_2\cdot A^2) \, \left[\Omega(m)\,
      S+C\right]
    \label{eq:order2}
    \,.
  \end{align}
Finally, setting $e=0$ gives the second-order correction to $S$,
from which the correction to the physical quark propagator $S_q$ is readily
obtained using~\cref{eq:ovprop},
\begin{equation}
  \tfrac{1}{2}\partial_e^2 S_q = [\mathcal{P}^{-1} \, \tfrac{1}{2}(\partial_e^2 S) \, \mathcal{R}_5
  \mathcal{P}]_{11}\,.
  \label{eq:order2_phi}
\end{equation}
From~\cref{eq:order2} we observe that when $C=0$ the term in the first line corresponds to a double
insertion of the electromagnetic vector current, while the second line corresponds to a single
second-order current insertion. 
On the other hand, when $C\neq 0$ other contributions appear, which cannot be interpreted
in terms of proper current insertions and arise from the operator $D_-$ in the definition
of the antiquark field. Writing the second derivative of the field $\bar{q}$ as
\begin{equation}
  \partial_e^2\bar{q} = (\bar{Q}_2\cdot A^2)\,,
\end{equation}
with 
\begin{equation}
  (\bar{Q}_2\cdot A^2)(x) = \sum_{\mu,z\in\Lambda} A_\mu(z)^2 \, \bar{Q}_2^\mu(z,x)\,,
\end{equation}
and 
\begin{equation}
  \bar{Q}_2^\mu(z,x) = -\sum_{x'\in\Lambda} \{\bar{\psi}(x')\,\Gamma_2^\mu(x',z,x)\,C\,\mathcal{R}_5\mathcal{P}\}_1 \,,
\end{equation}
we can rewrite the second-order correction to the physical quark propagator as 
\begin{widetext}
\begin{align}
  \tfrac{1}{2}\partial_e^2 S_q(x,y) &= - \, \sum_{\mu,\nu}\sum_{z,z'\in\Lambda} A_\mu(z)A_\nu(z') \, \Big\{
    \langle q(x) V^\nu(z') V^\mu(z) \bar{q}(y) \rangle - \langle q(x) V^\nu(z') \bar{Q}_1^\mu(z,y) \rangle \Big\}
  \nn\\
  & \quad - \frac{1}{2}\sum_{\mu,z\in\Lambda} A_\mu(z)^2 \, \Big\{
    \langle q(x) T^\mu(z) \bar{q}(y) \rangle + \langle q(x) \bar{Q}^\mu_2(z,y) \rangle
  \Big\}\,.
\end{align}
\end{widetext}
Here we have introduced the DWF seagull current (also called ``tadpole current'' in the literature)
\begin{equation}
  T^{\mu}(z)=\sum_{x,y\in\Lambda} \bar\psi(x) \, \Gamma_2^\mu(x,z,y) \, \Omega(m) \psi(y)\,,
  \label{eq:dwf_seagull_current}
\end{equation}
which can be explicitly written as
\begin{align}
  T^\mu(z) &=\frac{1}{2}[\bar{\psi}(z)(1-\gamma_{\mu})U_{\mu}(z)\Omega(m)\psi(z+\hat{\mu})\nn\\
  &\qquad+\bar{\psi}(z+\hat{\mu})(1+\gamma_{\mu})U_{\mu}^\dagger(z)\Omega(m)\psi(z)
  ]\,,
\end{align}
and is related by Ward identities to the conserved current $V^\mu(z)$, as will be shown below.

To visualize these expressions as Feynman diagrams, we need to introduce the following rules
for the seagull currents
\begin{align}
  \tcurrent &=
  (\Gamma_2\cdot A^2)\,\Omega(m)\,,\\
  \dtcurrent &=
  -(\Gamma_2\cdot A^2)\, C\,.
\end{align}
The second-order correction to $S$ in~\cref{eq:order2} can be written as
\begin{widetext}
  \begin{equation}
    S\underset{\mathcal{O}(e^2)}{=}e^2\Bigg(
      \eTwojj - \eTwojdj - \frac{1}{2}\, \eTwot + \frac{1}{2}\, \eTwodt\,\,
    \Bigg)+\mathcal{O}(e^3)\,.
  \end{equation}
\end{widetext}

\subsection{Ward-Takahashi identity for DWF currents}

After promoting the $\SU(3)$ gauge links to $\U(3)$ links in the definition of the
DWF conserved current in~\cref{eq:dwf_conserved_current}, an infinitesimal vector
transformation of the fermion fields gives
\begin{equation}
  \delta_V V^\mu(x) = -ie \, (\partial_\mu\omega)(x) \, T^\mu(x)\,.
\end{equation}
Using now the Ward-Takahashi identity derived in~\cref{eq:wti} with the operator
$O=V^\nu(0)$, we obtain the following identity for the domain-wall fermion currents
\begin{equation}
  {\textstyle\sum}_{\mu}\partial_{\mu,z}^* \braket{V^\mu(z)V^\nu(0)} =
   - \left[\delta(z)-\delta(z-\hat{\nu})\right] \braket{T^\nu(0)}\,,
\end{equation}
In momentum space this transforms into
\begin{equation}
  {\textstyle\sum}_{\mu} \, \hat{q}_\mu \, \Pi_{\mu\nu}(\hat{q}) = 0\,,
\end{equation}
where $\hat{q}_\mu = 2\sin(q_\mu/2)$ is the lattice momentum, and
$\Pi_{\mu\nu}(\hat{q})$ is the vacuum polarization tensor~\cite{Gockeler:2003cw}
\begin{align}
  \Pi_{\mu\nu}(\hat{q}) &= \sum_{x\in\Lambda} e^{iq(x+(\hat{\mu}-\hat{\nu})/2)} \langle
  V_\mu(x) V_\nu(0) \rangle + \delta_{\mu\nu} \langle
  T_\mu(0) \rangle\,.
\end{align}
This relation shows that the appearance of the seagull current at second order
in the expansion of the DWF propagator guarantees the cancellation of a contact term
arising when the position of the two conserved vector currents in correlation functions
coincide.

\section{Higher-order corrections}
\label{sec:orderN}
It is straightforward to extend the derivation of the first- and second-order
corrections to the quark propagator to arbitrary order $n$. The $n$-th derivative
in $e$ of the Wilson-Dirac operator is given by
\begin{align}
    e^n \partial_e^n D_\mathrm{w} &= - (-ie)^n \, (\Gamma_n \cdot A^n) \,, 
\end{align}
where
\begin{align}
    (\Gamma_n \cdot A^n)(x,y) &= \sum_{\mu,z\in\Lambda} A_\mu(z)^n \, \Gamma_n^\mu(x,z,y) \,,
\end{align}
and we have introduced the point-split operator appearing at $n$-th order as
\begin{align}
    \Gamma_n^\mu(x,z,y) &= \Gamma_\mu^-(x,y) \, \delta(z-y) \\
    &\qquad\qquad + (-1)^n \, \Gamma_\mu^+(x,y) \, \delta(z-x) \nn\\
    &=
    \begin{cases}
        \Gamma_1^\mu(x,z,y) \quad \text{for $n$ odd} \\
        \Gamma_2^\mu(x,z,y) \quad \text{for $n$ even} 
    \end{cases}
    .
\end{align}
As a consequence, the $n$-th derivative in $e$ of the generalized domain-wall Dirac operator reads
\begin{align}
    e^n \partial_e^n D_\mathrm{dw} &= - (-ie)^n \, (\Gamma_n \cdot A^n) \, \Omega(m)\,,
\end{align}
and one can use this result to derive the $n$-th derivative in $e$ of the physical quark propagator as
\begin{align}
    & \partial_e^n S = \sum_{k=0}^n\binom{n}{k} \,  [\partial_e^{k} S_\mathrm{dw}(m)] \, [\partial_e^{n-k} D_-] \,,\\
    & \partial_e^n S_q = [\mathcal{P}^{-1} \, (\partial_e^n S) \, \mathcal{R}_5 \mathcal{P}]_{11}\,.
\end{align}

\section{Quark number susceptibilities}
\label{sec:qnumber}
In this section we discuss how the results derived above in the context of electromagnetic
corrections can also be applied to the study of QCD thermodynamics at finite temperature,
in particular to the computation of quark number susceptibilities.

The introduction of a quark chemical potential, $\upmu$, in the domain-wall fermion
formalism has been discussed in Refs.~\cite{Bloch:2007xi,Hegde:2008nx,Banerjee:2008ii,Brower:2012vk}.
It amounts to a redefinition of the Wilson-Dirac operator $D_\mathrm{w}$ in~\cref{eq:DW},
where the $\SU(3)$ gauge links in the temporal direction are modified as
\begin{equation}
  U_4 \to e^{-\upmu} \, U_4\,, 
  \qquad
  U_4^\dagger \to e^{\upmu} \, U_4^\dagger\,.
\end{equation}
If the chemical potential is imaginary, this transformation can
be interpreted as a rephasing of the gauge links by a $\U(1)$ factor, analogous to~\cref{eq:u3links},
with the specific choice of background field $e A_\nu = i \upmu \, \delta_{\nu,4}$.

In QCD at finite temperature and density, the thermodynamic behavior of strongly
interacting matter is governed by fluctuations of conserved charges, such as baryon
number, electric charge, and strangeness. These fluctuations are encoded in quark number
susceptibilities, which are obtained as second derivatives of the QCD partition function
with respect to the quark chemical potentials.
Assigning distinct chemical potentials $\upmu_{f}$  to each flavor $f$, and considering a
theory with two light quarks and one strange quark, the diagonal and off-diagonal
susceptibilities are given by
\begin{align}
  \chi_2^{f} &= \frac{T}{V} \left. 
    \frac{\partial^{2}}{\partial \upmu_{f}^2}\, 
    \log \mathcal{Z}(T,V,\vec{\upmu})
    \right|_{\vec{\upmu}=0}\,,
  \\
  \chi_{11}^{f\!f'} &= \frac{T}{V} \left.
    \frac{\partial^{2}}{\partial \upmu_{f}\,\partial\upmu_{f'}} \, 
    \log \mathcal{Z}(T,V,\vec{\upmu})
    \right|_{\vec{\upmu}=0}\,,
\end{align}
where $f,f'\in\{u,d,s\}$ (with ${f'\neq f}$), $T$ denotes the temperature, $V$ the spatial volume, and
${\vec{\upmu}=\{\upmu_u,\upmu_d,\upmu_s\}}$  the set of quark chemical potentials. 

Following the strategy of Refs.~\cite{Bloch:2007xi,Goswami:2025euh},
we consider the QCD partition function formulated with overlap fermions.
After integrating out the fermion fields, the partition function is expressed in terms of the fermionic
determinant $\det[D_\mathrm{ov}(m)]$, which is related to the determinant of
the domain-wall fermion Dirac operator by~\cref{eq:ov_dwf_correspondence}, namely
\begin{equation}
  \det[D_\mathrm{ov}(m)] = \frac{\det[D_\mathrm{dw}(m)]}{\det[D_\mathrm{dw}(1)]}\;.
\end{equation}
Defining $D_{f} = D_\mathrm{ov}^{(f)}(m_f)$ the overlap Dirac operator for quarks of
flavor $f$ and mass~$m_f$, we use the relation
\begin{equation}
  \frac{\partial}{\partial\upmu_{f}} \log \det[D_{f}] = \tr
  \left(S_{f}\frac{\partial D_{f}}{\partial\upmu_{f}}\right)\,,
\end{equation}
with $S_{f}=D_{f}^{-1}$, to obtain the susceptibilities~\cite{Gottlieb:1987ac,Gottlieb:1988cq,Cheng:2009be}
\begin{align}
  \chi_2^{f} &= \frac{T}{V} \bigg\{
    {\bigg\langle
      \tr\!\bigg(S_{f} \frac{\partial^{2} D_{f}}{\partial\upmu_{f}^{2}}
        - S_{f} \frac{\partial D_{f}}{\partial\upmu_{f}} \,
          S_{f} \frac{\partial D_{f}}{\partial\upmu_{f}}\bigg)
    \bigg\rangle}_{\vec\upmu=0}
    \notag\\
    & \qquad +
    {\bigg\langle
      \tr\!\bigg(S_{f} \frac{\partial D_{f}}{\partial\upmu_{f}}\bigg)
      \tr\!\bigg(S_{f} \frac{\partial D_{f}}{\partial\upmu_{f}}\bigg)
    \bigg\rangle}_{\vec\upmu=0}
  \bigg\}\,,
  \\[8pt]
  \chi_{11}^{f\!f'} &= \frac{T}{V} \,
    {\bigg\langle
      \tr\!\bigg(S_{f} \frac{\partial D_{f}}{\partial\upmu_{f}}\bigg)
      \tr\!\bigg(S_{f'} \frac{\partial D_{f'}}{\partial\upmu_{f'}}\bigg)
    \bigg\rangle}_{\vec\upmu=0}\,.
\end{align}
It is easy to show that these traces can equivalently be written in terms of
domain-wall fermion operators as 
\begin{equation}
  \tr\!\big[F(D_f)\big] 
  = \tr\!\big[F(D_\mathrm{dw}^{(f)}(m_f))\big] 
  - \tr\!\big[F(D_\mathrm{dw}^{(f)}(1))\big]\,.
\end{equation}
The derivatives of the domain-wall Dirac operator with respect to the quark chemical
potential can then be obtained using the results of the previous sections, through the
analytic continuation $e A_\nu \to i\delta_{\nu,4}\upmu_f$, as
\begin{align}
  \frac{\partial}{\partial\upmu_{f}} D_\mathrm{dw}^{(f)}(m_f)
  &=  -\big[\Gamma_4^{-}-\Gamma_4^{+}\big] \, \Omega(m_f) \, ,\\
  \frac{\partial^{2} }{\partial\upmu_{f}^{2}} D_\mathrm{dw}^{(f)}(m_f)
  &= -\big[\Gamma_4^{-}+\Gamma_4^{+}\big] \, \Omega(m_f) \, .
\end{align}

Finally, using the results of~\cref{sec:orderN}, one can easily extend this
derivation to higher-order susceptibilities.

\section{Numerical implementation}
\label{sec:numerical}
\begin{figure*}[htb]
  \centering
  \includegraphics{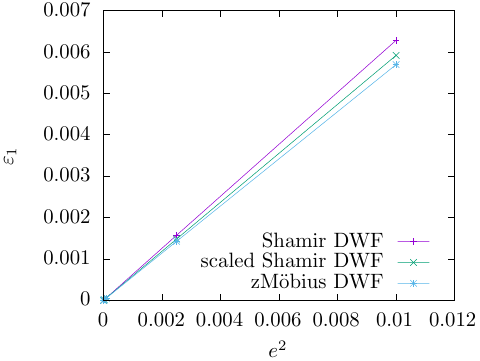}\includegraphics{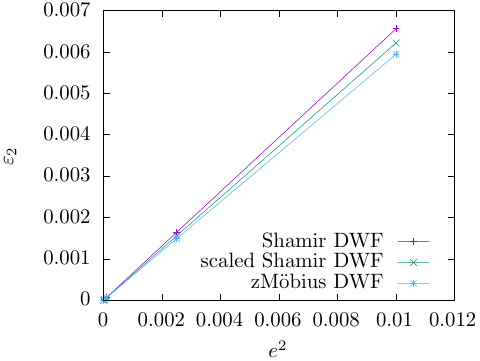}
  \caption{Representative example of relative errors between numerical derivatives of the
  quark propagator with respect to the electric charge and the perturbative expansions derived in this
  work. The left panel is a test of the first-order expansion \cref{eq:order1}, and the
  right panel is a test of the second-order expansion \cref{eq:order2}. In both cases, the
  horizontal axis parameterizes the squared electric charge, and relative errors are
  plotted for three different DWF actions. Points represent the actual data, and lines are
  simply joining the points. Further details of the calculation are provided in the text.}
  \label{fig:check}
\end{figure*}
We implemented numerically the first- and second-order currents described in \cref{sec:order1,sec:order2} using the Grid library (cf.~Refs.~\citep{Boyle:2015tjk,Yamaguchi:2022feu} and the repository \footnote{\url{https://github.com/paboyle/Grid}}). Our code is publicly released at~\citep{portelli_2026_21496822}, and in particular the currents routines are available as a single C++ header file that can be embedded in other Grid programs.

In the code provided with this work, we also implemented a validation test of the first- and second-order expansions \cref{eq:order1,eq:order2}, for different parameterizations of the DWF operator, as described below.
One can reparameterize the kernel function in \cref{eq:kernel_function} by writing
$b_s + c_s = (b + c)/\omega_s$ and keeping $b_s - c_s = b - c$ fixed. Then,
\begin{equation}
  H_s = \frac{1}{\omega_s} H_\mathrm{M}(b,c)\,,
\end{equation}
with
\begin{equation}
  H_\mathrm{M}(b,c) = \gamma_5 \, \frac{(b+c)\,D_\mathrm{w}}{2+(b-c)D_\mathrm{w}}
\end{equation}
denoting the M\"obius kernel~\cite{Brower:2004xi,Brower:2005qw,Brower:2012vk}. The
corresponding transfer matrix becomes
\begin{equation}
  \tm^{-L_s} = \prod_{s=1}^{L_s} \,
  \frac{\omega_s+H_\mathrm{M}(b,c)}{\omega_s-H_\mathrm{M}(b,c)}\,,
\end{equation}
providing the approximation to the sign function $\mathrm{\epsilon}_{L_s}[H_\mathrm{M}(b,c)]$, with $\mathrm{\epsilon}_{L_s}$ defined by
\begin{equation}
  \mathrm{\epsilon}_{L_s}(x) =
  \frac{\prod_s[\omega_s+x]-\prod_s[\omega_s-x]}
  {\prod_s[\omega_s+x]+\prod_s[\omega_s-x]}
  \,.
\end{equation}
Different choices of the parameters $\omega_s$, $b$, and $c$ define distinct kernel
functions, leading to different approximations of the overlap Dirac operator:
\begin{itemize}[leftmargin=*]
  \item $\omega_s = 1$, $b = 1$, $c = 1$ gives the Wilson kernel ${H_\mathrm{w} = \gamma_5
              D_\mathrm{w}}$, used in the Neuberger
        operator~\cite{Neuberger:1997bg,Neuberger:1997fp}.
  \item $\omega_s = 1$, $b = 1$, $c = 0$ gives the Shamir kernel $H_\mathrm{S} = \gamma_5
          D_\mathrm{w} / (2 + D_\mathrm{w})$~\cite{Furman:1994ky,Shamir:1993zy}.
  \item Keeping $b - c = 1$ and varying $b + c = \alpha$ defines the ``scaled Shamir''
        kernel, $H_{\mathrm{S}}^{(\alpha)}=\alpha H_\mathrm{S}$. For instance, $\alpha = 2$ is
        used in the RBC-UKQCD physical point ensembles~\cite{RBC:2014ntl}.
  \item In Refs.~\cite{Chiu:2002ir,Chiu:2002kj}, $\omega_s$ are tuned so that
        $\epsilon_{L_s}(\omega_s^{-1}H_\mathrm{S})$ becomes the Zolotarev approximation to the
        sign function.
  \item Finally, the so-called zM\"obius approximation~\cite{Mcglynn:2015uwh} is obtained
        by letting $\omega_s$ take on complex values and tuning the coefficients in such a way
        that $\epsilon_{L_s}(\omega_s^{-1}H_\mathrm{S})$ reproduces the approximate sign
        function obtained with a different kernel $\epsilon_{L_s'}(\alpha
          H_{\mathrm{S}})$ and a
        different value of $L_s$, with $L_s < L_s'$. This allows a reduction in the
        computational cost.
\end{itemize}

We now consider a fixed gauge configuration $U_{\mu}$ and an electromagnetic potential $A_{\mu}$, as well as a given physical four-dimensional source vector $\eta$ (i.e.~a four-dimensional field with one color and one spin index). The corrections in \cref{eq:order1,eq:order2} can then be constructed as follows, where $e=0$ is assumed in all equations.
\begin{enumerate}[leftmargin=*]
  \item Build a
        five-dimensional source $\eta_5$ from the four-dimensional one as a vector with components ${(\eta_5)_s = [\mathcal{R}_5\mathcal{P}]_{s1}\, \eta}$, namely
        \begin{equation}
          \eta_5=(P_+\eta, 0, \dots, 0, P_-\eta)^T\,;
        \end{equation}
  \item solve numerically the linear system
        \begin{equation}
          D_\mathrm{dw}(m)\phi=D_-\eta_5\,,
          \label{eq:solve_1}
        \end{equation}
        where $\phi$ is the solution vector. One then has ${\phi=S\,\eta_5}$, and the physical quark propagator is given by
        \begin{equation}
          S_q\,\eta=[\mathcal{P}^{-1}S\,\eta_5]_{11}\,,
        \end{equation}
        which is a direct application of \cref{eq:ovprop};
  \item build the sequential source
        vector
        \begin{equation}
          \eta'=-i(\Gw{A})\big[\Omega(m)\,\phi+C\,\eta_5 \big]\,;
        \end{equation}
  \item solve numerically the linear system
        \begin{equation}
          D_\mathrm{dw}(m)\phi'=\eta'\,,
          \label{eq:solve_2}
        \end{equation}
        where $\phi'$ is the new solution vector. One then has
        \begin{equation}
          \phi'=(\partial_eS\,\eta_5)|_{e=0}\,,
        \end{equation}
        which implements \cref{eq:order1};
  \item build the sequential source
        vector
        \begin{equation}
          \eta'' = -i (\Gamma_1\cdot A)\Omega(m)\,\phi'
          - \tfrac{1}{2}(\Gamma_2\cdot A^2)\big[\Omega(m)\,\phi + C\,\eta_5
            \big]\,;
        \end{equation}
  \item solve numerically the linear system
        \begin{equation}
          D_\mathrm{dw}(m)\phi''=\eta''\,,
        \end{equation}
        where $\phi''$ is the solution vector. One then has
        \begin{equation}
          \phi''=\tfrac{1}{2}(\partial^2_eS\,\eta_5)|_{e=0}\,,
        \end{equation}
        which implements \cref{eq:order2}.
\end{enumerate}

In order to check the procedure above, one needs a numerical ground truth for comparison. Such reference can be obtained by computing numerically the derivatives $(\partial_eS\,\eta_5)|_{e=0}$ and $(\partial^2_eS\,\eta_5)|_{e=0}$. For the sake of simplicity, we will only consider the components of the currents $\Gamma_1$ and $\Gamma_2$ in an arbitrary direction $\mu_0$, which can be achieved by assuming the electromagnetic potential is given by the constant $A_{\mu}(x)=\delta_{\mu\mu_0}$. We denote by $\mathcal{U}_{\mu}(e)$ the associated $\U(3)$ links defined in \cref{eq:u3links} for charge $e$.
We additionally define $\phi(e)$ as the solution of \cref{eq:solve_1} where the gauge links are set to $\mathcal{U}_{\mu}(e)$ in both the $D_\mathrm{dw}(m)$ and $D_-$ operators. For a given $e>0$, one can compute $\phi(e)$, $\phi(-e)$, and $\phi(0)$, and then form the following finite differences
\begin{align}
  \delta_1\phi & =\frac{1}{2e}[\phi(e)-\phi(-e)]\,,            \\
  \delta_2\phi & =\frac{1}{2e^2}[\phi(e)+\phi(-e)-2\phi(0)]\,,
\end{align}
which for small $e$ have the expansions
\begin{align}
  \delta_1\phi & =(\partial_eS\,\eta_5)|_{e=0}+\mathcal{O}(e^2)\,, \\
  \delta_2\phi & =\tfrac{1}{2}(\partial^2_eS\,\eta_5)|_{e=0}
  +\mathcal{O}(e^2)\,.
\end{align}
By varying $e$, we can then check that the relative errors
\begin{align}
  \epsilon_1 & =2\frac{\|\delta_1\phi-\phi'\|}{\|\delta_1\phi+\phi'\|}\,,   \quad
  \epsilon_2  =2\frac{\|\delta_2\phi-\phi''\|}{\|\delta_2\phi+\phi''\|}\,,
\end{align}
indeed scale linearly with $e^2$. We show in \cref{fig:check} the quantities $\epsilon_1$ and $\epsilon_2$ computed on an $8^4$ lattice, with a fixed random $\SU(3)$ gauge field $U_{\mu}$ and a fixed random source $\eta$.
We compute these relative error terms for three different choices of the DWF operator, all using ${L_s=10}$. The first choice is the Shamir operator, the second choice is the $\alpha=2$ scaled Shamir operator, and finally the third choice is the zMöbius operator with the same parameters as those used in e.g.\ Refs.~\cite{Boyle:2022lsi,Hodgson:2023}. The DWF operators are inverted using a red-black preconditioned conjugate gradient algorithm with a residual of $10^{-16}$, which is conservatively chosen to remain negligible for the numerical derivatives $\delta_1\phi$ and $\delta_2\phi$ and our choice of values for the charge $e$. In all cases, we can observe that the errors $\epsilon_1$ and $\epsilon_2$ scale linearly with $e^2$ and vanish as $e\to0$, validating numerically our derivation of \cref{eq:order1,eq:order2}. Full logs of the runs leading to these results are provided in the repository~\citep{portelli_2026_21496822}.

\section{Conclusion}
\label{sec:conclusion}

In this work, we have presented a comprehensive and rigorous derivation of the perturbative expansion of the generalized DWF Dirac operator under electromagnetic corrections up to $\mathcal{O}(e^2)$. This theoretical framework is important for providing fully gauge-invariant lattice calculations targeting isospin-breaking effects and radiative corrections to hadronic processes involving chiral fermions.

By mapping the explicit dependence of the four-dimensional physical quark fields to the underlying five-dimensional fields via the (QCD+QED) gauge links in the $D_-$ operator, we demonstrated how expanding the correlation functions with respect to the electric charge generates additional local contact terms. Employing a background-field method, we successfully re-derived the established first-order $\mathcal{O}(e)$ corrections and, crucially, obtained the previously missing second-order $\mathcal{O}(e^2)$ local operator insertions, corresponding to the so-called seagull vertices, which are strictly necessary to maintain gauge invariance and full theoretical consistency.

We have additionally evaluated the consequences of this formulation on the lattice Ward-Takahashi identities, explicitly tracking the modifications imposed by these link-dependent field definitions. Furthermore, to show the versatility of our derivation, we applied it to the domain of QCD thermodynamics. Specifically, we outlined the corrections needed to compute quark number susceptibilities via the analytic continuation of a flavor chemical potential ($eA_4 \rightarrow i\upmu_f$).

For practical applications, the entire mathematical setup has been implemented and validated within the Grid data-parallel C++ library framework, ensuring that these new vertices can be deployed directly in high-performance computing environments. The corresponding code is publicly available at~\citep{portelli_2026_21496822}.

Looking forward, this complete set of $\mathcal{O}(e^2)$ vertices provides the theoretical foundation required to execute fully consistent ab initio computations of $\mathcal{O}(\alpha_{\mathrm{em}})$ corrections directly at the physical point using generalized domain-wall fermions. By properly accounting for these contact terms, future lattice studies of radiative corrections to leptonic and semi-leptonic hadronic decays will have a highly controlled avenue to eliminate systematic errors, ultimately paving the way for more stringent precision tests of CKM unitarity.

\vfill
\begin{acknowledgments}
  The authors would like to thank Peter A.~Boyle and the members of the
  RBC-UKQCD collaboration for useful discussions. M.D.C.~and A.P.~have been supported
  in part by UK STFC grants ST/P000363/1, and A.P.~additionally by grant ST/X000494/1. A.P. also received funding from the
  European Research Council (ERC) under the European Union's Horizon 2020
  research and innovation programme under grant agreements No 757646 \& 813942.
  M.D.C. has received funding from the European Union's Horizon Europe research
  and innovation programme under the Marie Sklodowska-Curie grant agreement No. 101108006.
\end{acknowledgments}
\bibliography{gdwf-qed}
\end{document}